\documentclass[num-refs]{wiley-article}

\usepackage{siunitx}
\usepackage{booktabs}
\usepackage[utf8]{inputenc}
\usepackage{hyperref}
\DeclareUnicodeCharacter{03A8}{\ensuremath{\Psi}}
\DeclareUnicodeCharacter{2212}{\ensuremath{-}}

\papertype{Review Article}
\title{Spatial statistics for screening molecular structures}

\author[1,2,3]{Pranoy Ray}%\authfn{1}
\author[1,2]{Surya R. Kalidindi}%\authfn{2}

\affil[1]{George W. Woodruff School of Mechanical Engineering, Georgia Institute of Technology, Atlanta, GA 30332}
\affil[2]{School of Computational Science and Engineering, Georgia Institute of Technology, Atlanta, GA 30332}
\affil[3]{Department of Chemistry, The University of Chicago, Chicago, IL 60637}

\corraddress{Dr. Surya R. Kalidindi, George W. Woodruff School of Mechanical Engineering, Georgia Institute of Technology, Atlanta, GA 30332}
\corremail{surya.kalidindi@me.gatech.edu}

\fundinginfo{NSF DMREF Award 2119640}

\runningauthor{Ray et al.}

\begin{document}

\begin{frontmatter}
\maketitle

\begin{abstract}
The dominant paradigm in computational materials discovery relies on heavily parameterized deep architectures, including message-passing graph networks and equivariant models, that typically require large DFT-labeled training sets and produce non-convex latent representations that complicate continuous optimization for inverse design. These architectures are impractical in data-scarce regimes, which is the typical case in molecular screening, and exhibit well-documented limitations in capturing chemically disordered configurations and chiral geometries. This review presents feature engineering based on spatial statistics as a physically rigorous and immediately deployable alternative. Molecular structures are encoded as voxelized scalar fields, and two-point auto- and cross-correlations are evaluated deterministically via Fast Fourier Transforms, transferring spatial pattern recognition from the learning algorithm to a closed-form, physics-informed operation. Principal component analysis of the resulting correlation maps yields low-dimensional, convex representations that support lean neural networks ($<$100k trainable parameters) and non-parametric surrogate models, achieving sub-2\% normalized prediction error from as little as 0.1\% of the dataset in specific reported cases. Demonstrated across periodic crystals, chemically disordered high-entropy alloys, and non-periodic organic molecules, this framework enables Bayesian active learning and zero-shot extrapolation on commodity hardware, which remains challenging for current large-scale architectures at equivalent data budgets.
% Please include a maximum of seven keywords
\end{abstract}
\end{frontmatter}

\begin{figure*}[htbp]
    \centering
    \includegraphics[width=0.9\linewidth]{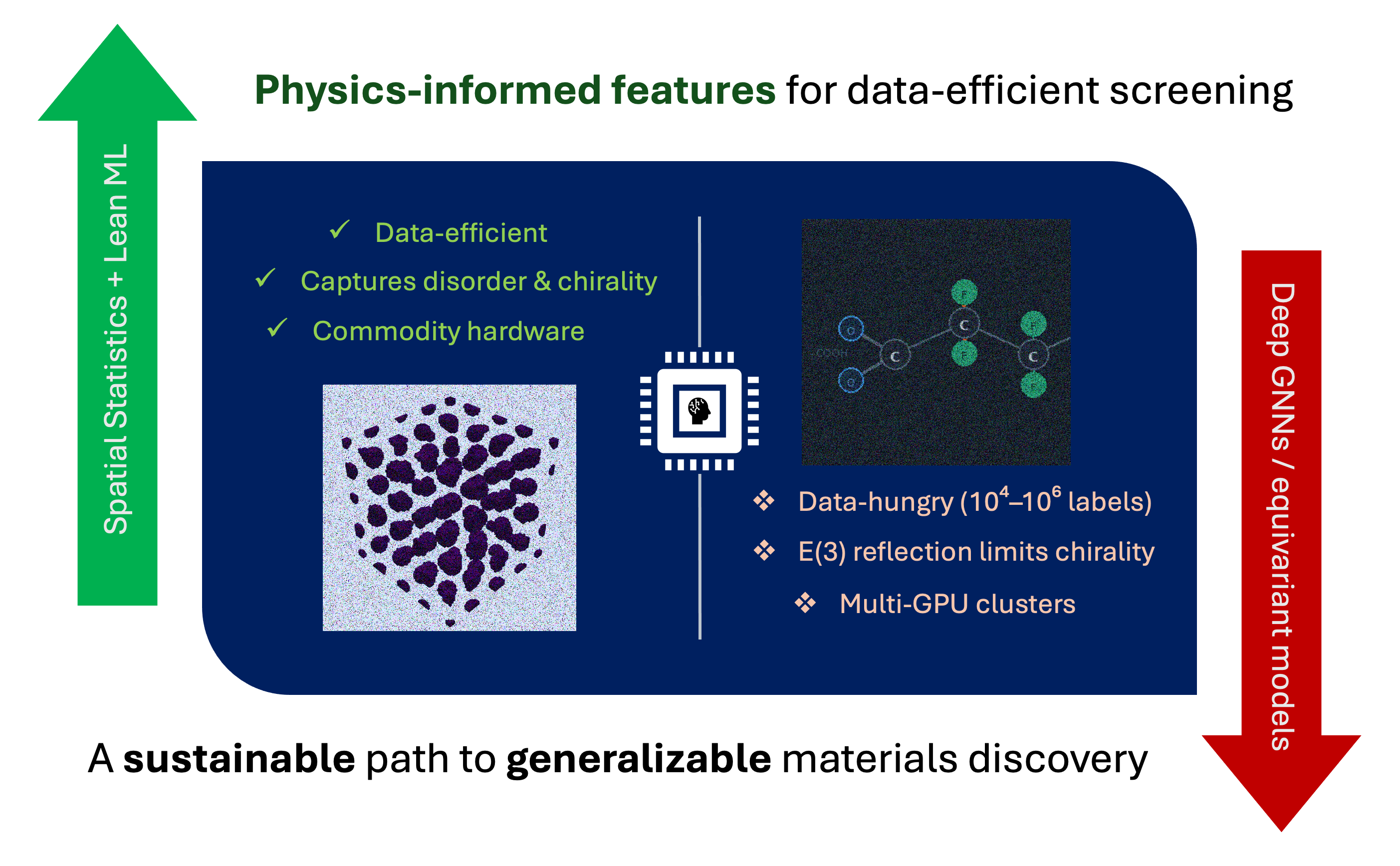}%graphical_abstract_spatial_feature_engineering
    \caption*{Graphical Abstract}
    \label{fig:graphical}
\end{figure*}

\section{Introduction}
The accelerated discovery and design of novel materials and complex molecular structures increasingly rely on data-driven surrogate models \cite{ramprasad_machine_2017, hastings_accelerated_2025, ward_general-purpose_2016, tabor_accelerating_2018, potyrailo_combinatorial_2011, ward_matminer_2018} to bypass the prohibitive computational costs of high-fidelity ab initio methods, such as Density Functional Theory (DFT) \cite{hohenberg_inhomogeneous_1964, kohn_self-consistent_1965, chakraborty_high_2021, nair_ti-decorated_2023, kundu_zr_2024, deb_copper_2023}. While contemporary deep learning architectures \cite{schutt_schnet_2018,deng_chgnet_2023} have demonstrated remarkable success in establishing re-usable process-structure-property linkages, their heavily parameterized nature demands both immense volumes of training data and places an excessive burden on the network architecture itself to implicitly learn spatial patterns: a task that scales poorly with structural complexity. This reliance creates a fundamental bottleneck in novel chemical spaces where data generation involves expensive self-consistent field (SCF) calculations \cite{wang_never-ending_2017, raza_machine_2019, pilania_accelerating_2013, curtarolo_high-throughput_2013, pilania_machine_2016}. Two divergent responses have emerged: a push toward foundation models pre-trained on tens of millions ($\geq{10^4}$) of DFT labels \cite{merchant_scaling_2023, levine_open_2026}, and a parallel pursuit of highly sample-efficient frameworks  \cite{hastings_accelerated_2025, qian_knowledge-driven_2023, ray_electronic_2026, barry_data-driven_2026} designed for rapid exploration of novel compositional and conformational spaces. Critically, no single foundation model yet generalizes with equal fidelity across structurally disparate classes; organic polymers, inorganic metallic alloys, and small molecules remain governed by fundamentally different physics, making system-agnostic, data-efficient representations an enduring necessity. 

\begin{figure}[h]
    \centering
    \includegraphics[width=\linewidth]{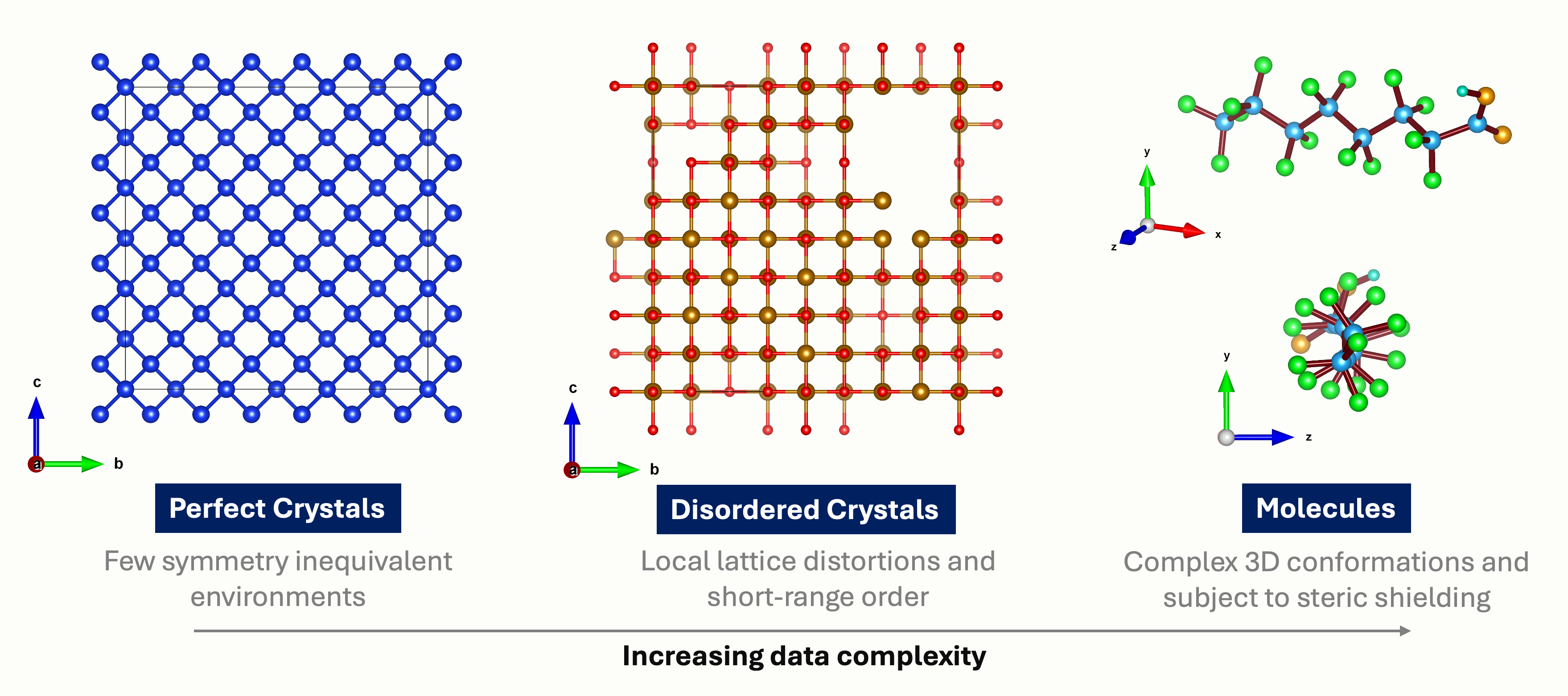}
    \caption{Hierarchical structural complexity across material classes. Schematic illustrating (left - Si) perfectly periodic crystals with a small number of symmetry-inequivalent environments, (center - Wüstite: $Fe_{1-x}O$) chemically disordered solid solutions with local lattice distortions and short-range order, and (right - PFOA) finite, non-periodic molecules with complex 3D conformations (helicity) and steric shielding (F atoms).}
    \label{fig:complexity}
\end{figure}

To navigate these fundamental data bottlenecks, the computational chemistry and physics communities have seen rapid and highly successful advancements in Machine Learned Interatomic Potentials (MLIPs)~\cite{bartok_gaussian_2010, bartok_representing_2013, batatia_mace_2022} and mathematically rigorous atomic representations, most notably the Smooth Overlap of Atomic Positions (SOAP)~\cite{bartok_representing_2013, bartok_machine-learning_2013}. MLIP architectures such as Gaussian Approximation Potentials (GAP) \cite{bartok_gaussian_2010} and MACE \cite{maes_mace_2025, batatia_mace_2022} excel at mapping local chemical environments to accurate Potential Energy Surfaces (PES) \cite{manzhos_neural_2021} for long-duration molecular dynamics. Similarly, descriptors such as SOAP \cite{bartok_representing_2013, bartok_machine-learning_2013} mathematically guarantee rotational, translational, and permutational invariance through the expansion of local atomic densities using spherical harmonics and radial basis functions. While these frameworks are exceptional for traversing local energy landscapes, they face inherent limitations in global property screening. Specifically, atom-centered descriptors such as SOAP often suffer from a combinatorial explosion in dimensionality as the number of constituent chemical species increases, limiting their rapid deployment in highly disordered, multi-component spaces. Furthermore, deploying highly parameterized MLIPs to predict global, volume-averaged properties represents an indirect and computationally over-parameterized route. Consequently, there remains a critical need for structural representations \cite{faber_crystal_2015, kajita_universal_2017, isayev_universal_2017, huang_communication_2016, kaundinya_machine_2021} that are simultaneously continuous, computationally lean, and globally invariant without suffering from species-based combinatorial scaling. Recent efforts have demonstrated the efficacy of transforming atomic coordinates into mathematically discrete, three-dimensional voxelized fields \cite{kaundinya_machine_2021, barry_voxelized_2020, ray_ml_2026, ray_electronic_2026}, offering a high-fidelity mapping of complex atomic neighborhoods that are otherwise difficult to capture using standard graph-based descriptors \cite{kaundinya_prediction_2022, choudhary_atomistic_2021}. As illustrated in Figure~\ref{fig:complexity}, the structural complexity of real materials spans three qualitatively distinct regimes: perfectly periodic crystals, chemically disordered solid solutions with broken translational symmetry and local lattice distortions, and finite non-periodic molecules. Graph-based architectures build a fixed-radius neighbor graph from a single set of atomic positions, which raises two difficulties in the disordered regime. First, connectivity-only (topological) graph descriptors discard the 3D geometric information that distinguishes conformers, and even geometric message-passing networks are provably bounded by the 1-Weisfeiler-Lehman test, so structures with locally identical environments can be indistinguishable \cite{xu_how_2018}. Second, a graph over one snapshot encodes a single realization of the disorder rather than its statistics. Two-point spatial correlations instead average over the field and directly quantify the co-occurrence of structural states at every vector separation, so different realizations at the same composition are separated by construction. Graph models can still encode disorder when the representation and sampling protocol are designed for it; the distinction is one of default suitability, not an absolute limitation. In contrast, voxelized fields encode the full spatial distribution of atomic species and/or the associated charge density, providing a natural, rigorous first-order statistical descriptor for disordered configurations across all three regimes in Figure~\ref{fig:complexity}.

The critical challenge of extracting low-dimensional regressors from these high-dimensional voxelized fields is addressed through the application of spatial statistics\cite{montes_de_oca_zapiain_convolutional_2021, robertson_micro2d_2024, generale_inverse_2024, kalidindi_application_2015, kalidindi_hierarchical_2015, harrington_application_2022}. Crucially, the mathematical formalisms\cite{fast_formulation_2011, kalidindi_hierarchical_2015} governing two-point spatial correlations are inherently scale-agnostic, excelling not only at the molecular level but also in quantifying complex disorder across the mesoscale\cite{mann_development_2022, harrington_application_2022, montes_de_oca_zapiain_convolutional_2021}. By mathematically encoding the translationally invariant topological patterns native to a spatial field, this deterministic feature-engineering step\cite{bostanabad_computational_2018, fullwood_gradient-based_2008, fast_formulation_2011} shifts the primary burden of spatial pattern recognition away from the learning algorithm itself\cite{mann_development_2022, ray_lean_2025}. In parallel, voxelized representations have also enabled multiscale localization frameworks such as recurrent localization networks\cite{kelly_recurrent_2021} and thermodynamically informed neural operators\cite{kelly_thermodynamically-informed_2025}, that operate directly on raw coefficient fields to predict full-field elastic strain evolutions. Localization problems require exact preservation of absolute spatial phase, which is intentionally averaged out in translationally invariant two-point correlations. Together, these homogenization- and localization-focused efforts highlight the versatility of discrete grid representations: the same voxelized fields can be routed either through correlation-based feature pipelines for property prediction, or through phase-preserving localization networks when the underlying physics demands it. Overall, spatial statistics and voxelized representations provide an interpretable, mathematically rigorous alternative to the opaque feature extraction processes embedded within deep convolutional stacks.

The explicit integration of engineered spatial features profoundly alters the architecture and data requirements of the corresponding predictive models \cite{kaundinya_machine_2021, gomberg_extracting_2017}. Operating directly on these reduced-order spatial manifolds enables the deployment of non-parametric models \cite{barry_voxelized_2023, ray_electronic_2026} and lean neural architectures \cite{ray_lean_2025, mann_development_2022} that eliminate dense, fully connected layers, thereby reducing the trainable parameter count by one to two orders of magnitude while maintaining high predictive accuracy. Furthermore, pairing PCA-reduced spatial correlations \cite{kalidindi_hierarchical_2015, fast_formulation_2011, kalidindi_novel_1970, kalidindi_feature_2020} with Gaussian Process Regression (GPR) \cite{rasmussen_gaussian_2006} naturally yields robust uncertainty quantification, making the framework ideal for Bayesian active learning \cite{lookman_active_2019, khatamsaz_bayesian_2023, balachandran_adaptive_2016, jablonka_bias_2021, hou_bayesian_2020, ray_refining_2025, ray_assessing_2026, stach_autonomous_2021, deneault_toward_2021, talapatra_autonomous_2018, ray_electronic_2026}. Such spatial-correlations-based feature engineering pipelines \cite{barry_voxelized_2023, ray_electronic_2026, ray_lean_2025} have achieved remarkable data efficiency, demonstrating less than 2\% normalized errors using merely 0.1\% of the training samples in certain cases \cite{barry_voxelized_2023, ray_electronic_2026}. Beyond crystalline solids, these spatial statistical descriptors have also shown to successfully isolate global properties dictated by 3D electron distributions, such as polarizability and thermodynamic stability in complex per- and polyfluoroalkyl substances (PFAS) \cite{ray_ml_2026}, underscoring their applicability across chemically diverse domains.

In parallel, a large body of work has focused on increasingly expressive deep architectures for atomistic learning, including message-passing graph neural networks \cite{choudhary_atomistic_2021, kaundinya_prediction_2022, xie_crystal_2018}, equivariant graph networks \cite{batzner_e3-equivariant_2022, dumitrescu_e3-equivariant_2025}, and 3D CNNs operating directly on voxelized charge densities \cite{zhao_predicting_2020, casey_prediction_2020, ray_lean_2025} or atomic neighborhoods \cite{xie_crystal_2018, schutt_schnet_2018, merchant_scaling_2023, choudhary_atomistic_2021, zhao_predicting_2020, casey_prediction_2020, shi_review_2024, gong_graph-based_2021, jiang_could_2020, yang_establishing_2019, singh_revealing_2026, maes_mace_2025, batatia_mace_2022, batzner_e3-equivariant_2022}. While these models can achieve high accuracy on benchmark datasets\cite{choudhary_joint_2020, levine_open_2026}, they typically require $10^5$–$10^7$ trainable parameters and DFT-generated training sets in the $10^4$–$10^6$ range to avoid overfitting, fundamentally limiting their practical deployment in novel data-scarce chemical spaces \cite{lei_universal_2022, barry_voxelized_2023, ray_lean_2025, kaundinya_machine_2021, kalidindi_hierarchical_2015}, when trained from scratch; pre-training, transfer, $\Delta$-, and multi-fidelity learning can substantially lower these requirements \cite{ramakrishnan_big_2015, smith_approaching_2019}. Moreover, the learned latent representations in such networks do not generally form convex, interpretable, or chemically agnostic manifolds, which can complicate gradient-based inverse design and Bayesian path-planning, in contrast to reduced-order spatial correlation–PCA feature spaces \cite{kalidindi_hierarchical_2015, barry_voxelized_2023, ray_electronic_2026, ray_ml_2026}.

This review synthesizes the theoretical foundations and demonstrated practical implementations of spatial feature engineering for molecular materials' screening. The subsequent sections will first detail the mathematical framework governing the generation of voxelized molecular structure fields and their quantification via spatial correlations. Following this, the discussion will contrast traditional deep learning architectures with lean models and Bayesian frameworks capable of exploiting these low-dimensional spatial features. Finally, a series of application case studies ranging from a zero-shot extrapolative discovery of high-entropy alloys to the conformational analysis of environmental contaminants will be presented, demonstrating the broad utility of this paradigm. The central argument is that, in data-scarce and extrapolative regimes, physics-informed spatial representations are a competitive, resource-light route to generalizable materials informatics, complementary to larger models and datasets. Within this scope, prior work can be broadly grouped into: (i) descriptor-centric approaches \cite{ward_general-purpose_2016, ward_matminer_2018} based on hand-crafted structural and compositional features; (ii) end-to-end deep learning models, including graph neural networks \cite{choudhary_atomistic_2021, xie_crystal_2018, schutt_schnet_2018} and 3D CNNs operating directly on atomistic graphs or charge-density fields; and (iii) physics-capturing spatial statistics \cite{kaundinya_machine_2021,barry_voxelized_2023,kalidindi_hierarchical_2015} that explicitly compute spatial correlations prior to lean model training. This review focuses primarily on the third category while critically comparing its capabilities and limitations with those of deep graph-based architectures. Table~\ref{tab:taxonomy} situates these representation families by conceptual origin, independent validation, and a representative application.

\begin{table}[h]
    \centering
    \footnotesize
    \caption{Taxonomy of structural representation families, separating conceptual origin, independent validation, and a representative application.}
    \label{tab:taxonomy}
    \begin{tabular}{p{2.4cm}p{3.2cm}p{3.2cm}p{3.2cm}}
        \toprule
        \textbf{Representation family} & \textbf{Conceptual origin} & \textbf{Independent validation} & \textbf{Representative application} \\
        \midrule
        Two-point spatial statistics & MKS / $n$-point correlations \cite{kalidindi_hierarchical_2015, fullwood_microstructure_2008} & Degeneracy \& realizability studies \cite{gommes_microstructural_2012, cluff_quantifying_2026}; versatile FFT algorithms \cite{cecen_versatile_2016} & HEA moduli, PFAS polarizability \cite{barry_voxelized_2023, ray_ml_2026} \\
        SOAP / ACE & Local density expansion \cite{bartok_representing_2013, drautz_atomic_2019} & GAP, MACE potentials \cite{bartok_gaussian_2010, batatia_mace_2022} & Interatomic potentials / PES \cite{manzhos_neural_2021} \\
        Wavelet scattering & Solid-harmonic scattering \cite{eickenberg_solid_2017} & Reused across cosmology and turbulence & QM9 molecular energies \cite{eickenberg_solid_2017} \\
        Graph / equivariant NNs & CGCNN, NequIP \cite{xie_crystal_2018, batzner_e3-equivariant_2022} & Matbench, OMol25 benchmarks \cite{levine_open_2026} & Formation energy, forces \cite{choudhary_atomistic_2021} \\
        Voxel 3D CNN & 3D voxel descriptors \cite{kajita_universal_2017, zhao_predicting_2020} & Multiple independent groups \cite{casey_prediction_2020} & Elastic properties from charge density \cite{ray_lean_2025} \\
        \bottomrule
    \end{tabular}
\end{table}

\section{Computation of spatial statistics for molecules}

\begin{figure}
    \centering
    \includegraphics[width=0.7\linewidth]{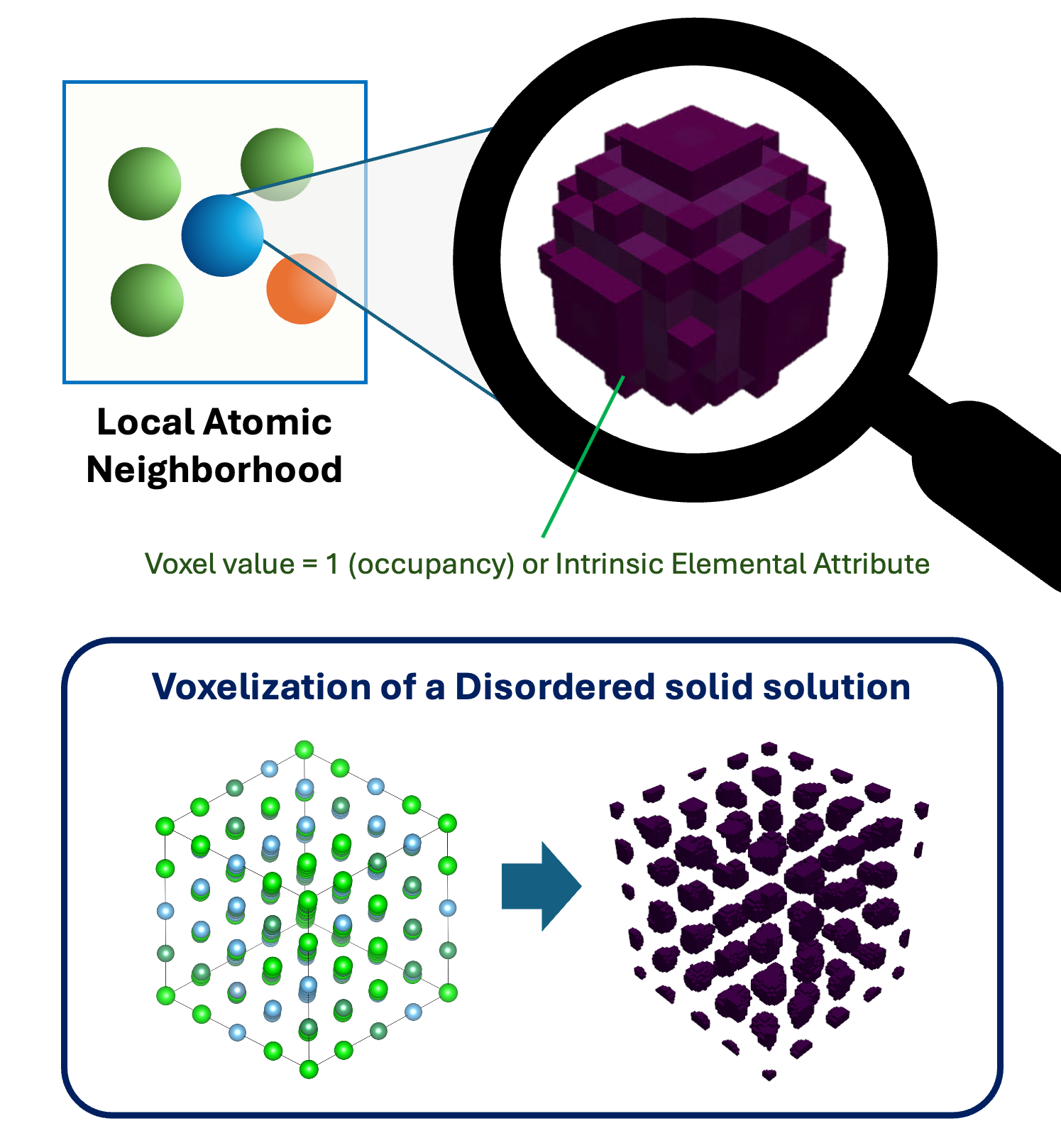}
    \caption{Voxelization pipeline for continuous atomic environments. The discrete, coordinate-based point-cloud of a local atomic neighborhood is systematically mapped onto a standardized cubic voxel grid. During this transformation, each voxel channel encodes specific physical attributes, such as binary occupancy, intrinsic elemental properties, or continuous pseudo-charge densities via a defined smearing function. Voxels falling within a prescribed atomic radius capture the designated elemental values, while interstitial regions remain at zero, effectively transforming the discrete Lagrangian coordinates into a continuous, multi-channel field ready for deterministic spatial statistical quantification.}
    \label{fig:voxel}
\end{figure}

\begin{figure}[h]
    \centering
    \includegraphics[width=0.7\linewidth]{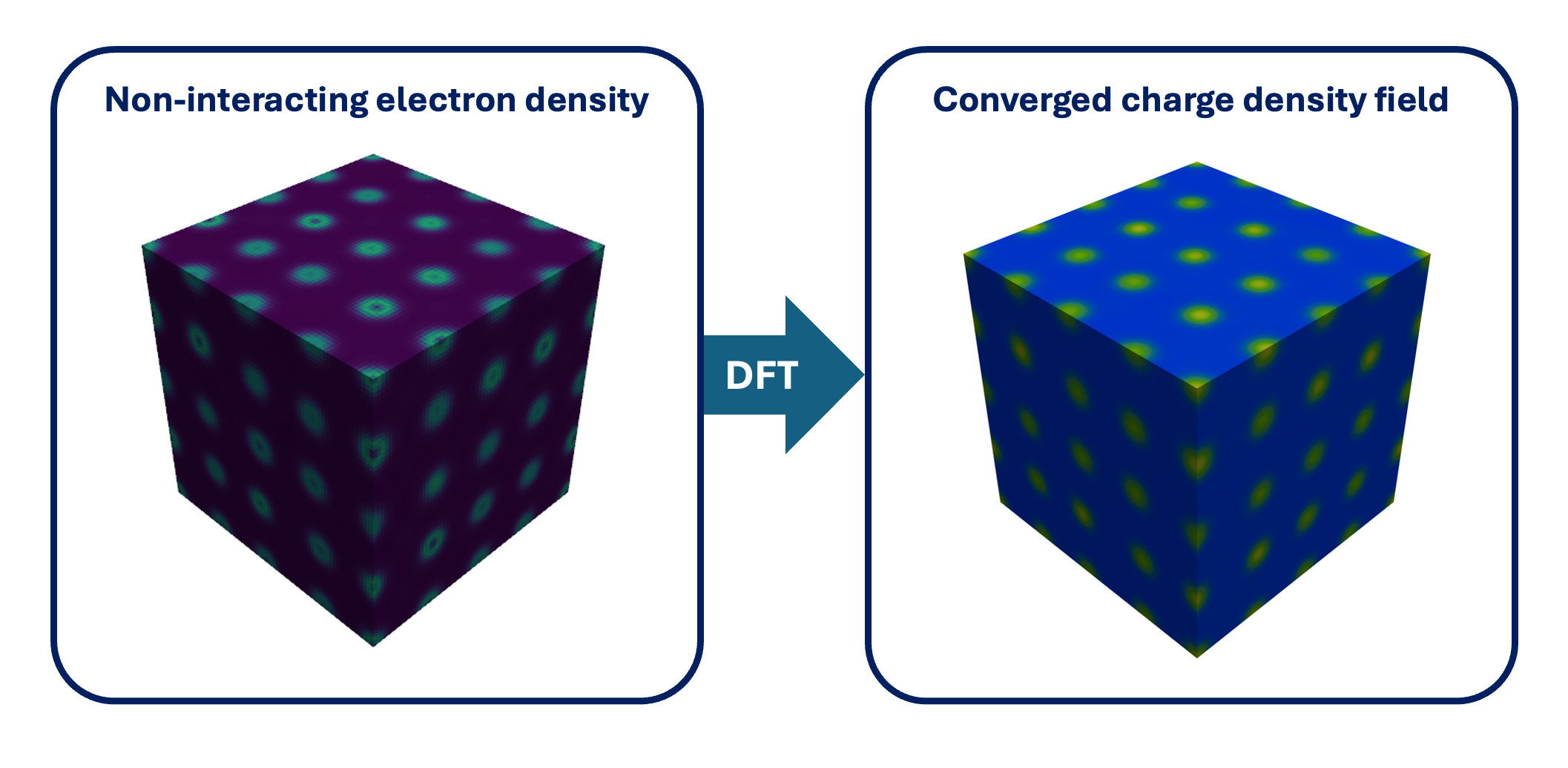}
    \caption{Sources of voxelized molecular structure data from DFT. Example shown for a refractory high-entropy alloy Al$_4$Nb$_{44}$Ti$_4$Zr$_{76}$, where the converged or pseudo charge density field is sampled on a regular grid to construct the voxelized representation used in the spatial feature engineering workflow.}
    \label{fig:voxelized_DFT}
\end{figure}

\subsection{Mathematical Infrastructure}
The fundamental prerequisite for spatial feature engineering lies in transforming coordinate-based atomic position representations into mathematically discrete three-dimensional voxelized fields (see Figure~\ref{fig:voxel}) that reflect accurately the spatial neighborhood details in the  molecular structure. This grid-based parameterization allows for the explicit embedding of complex topologies in physical fields, such as those quantified in quantum mechanics by the electron charge density (ECD); this representation can natively encode details of both the spatial arrangements and the nuanced chemical bonding environments of the constituent atoms \cite{casey_prediction_2020, zhao_predicting_2020, ray_lean_2025, barry_voxelized_2023, ray_electronic_2026}. Note that the voxelization protocols discussed here are applicable to virtually any type of spatial field. As examples, these may include converged charge densities (see Figure~\ref{fig:voxelized_DFT}) derived from density functional theory providing a rigorous quantum mechanical baseline or those constructed alternately by simply superimposing non-interacting pseudo-densities\cite{hamann_optimized_2013, hamann_norm-conserving_1979, hamann_generalized_1989} to avoid the computationally expensive self-consistent field (SCF) iterations \cite{lei_universal_2022,ray_electronic_2026}.

\subsubsection{Voxelization and Discrete Field Representations}

Mathematically, the atomic structure is mapped onto a discrete three-dimensional grid with voxel indices $S$, and a family of eigen–microstructure channels $\{\rho^{(m)}_s\}$ that each encode a single scalar attribute ($m$) on this grid\cite{kaundinya_machine_2021, ray_ml_2026, kalidindi_hierarchical_2015}. In the simplest binary occupancy limit, a voxel $s$ lying within a prescribed atomic sphere around any nucleus is assigned $\rho^{(0)}_s = 1$, and $\rho^{(0)}_s = 0$ otherwise, thereby reducing the molecular structure representation to a crisp indicator field. For richer, chemistry-aware representations, the binary representation can be replaced by intrinsic elemental attributes, such as electronegativity, heat of fusion, first ionization energy, or any other descriptor, while retaining zeros in the interstitial regions\cite{kaundinya_machine_2021, ray_ml_2026}. This construction effectively treats each attribute channel as an eigen microstructure in the formalism of the materials knowledge systems (MKS) framework\cite{kalidindi_hierarchical_2015}.

However, there is no universally accepted, exhaustive list of elemental attributes, and naive feature lists are known to introduce bias, incompleteness, and poor extrapolation beyond the training chemistry \cite{ward_general-purpose_2016, kaundinya_machine_2021}. A significantly more physics-grounded alternative is to work with pseudo charge density fields, $\tilde{\rho}_s$, obtained by superimposing non-interacting valence electron densities from norm-conserving pseudopotentials on the same voxel grid \cite{lei_universal_2022, ray_electronic_2026}. The physical efficacy of these pseudo-densities lies in their native ability to encode the spatial arrangement, atomic volume, baseline steric hindrance, and fundamental crystallographic symmetry of the system without requiring iterative quantum mechanical solving. The autocorrelations of this pseudo-density effectively capture the system's ``electronic packing manifold'' \cite{ray_electronic_2026}. While non-interacting pseudo-densities fundamentally lack information regarding charge transfer, electron correlation, and polarization (phenomena that only emerge during SCF convergence), the computation of cross-correlations between these pseudo-densities and intrinsic elemental attributes (such as electronegativity or first ionization energy) synthetically recovers first-order approximations of these complex chemical bonding environments \cite{kaundinya_machine_2021, ray_electronic_2026}. These mathematically engineered fields have been shown to form electronic manifolds that extrapolate across the tested alloy classes \cite{lei_universal_2022,ray_electronic_2026,barry_voxelized_2023}. In practice, we therefore consider a set of channels $\{\rho^{(m)}_s\}$ that includes occupancy, elemental attributes, and one or more pseudo-density fields, all expressed on a standardized real-space grid.

\subsubsection{Statistical Quantification and Dimensionality Reduction}
Once the molecular structure is represented as a discrete voxelized field, the extraction of its intrinsic topological and electronic regularities is achieved through rigorous spatial statistics. Specifically, the framework employs directionally resolved two-point auto- and cross-correlations to quantify the probability of finding specific structural states or field intensities at defined vector separations, and thus encode the translationally invariant topology of the field \cite{fast_formulation_2011, niezgoda_understanding_2011, fullwood_microstructure_2010, niezgoda_delineation_2008, fullwood_microstructure_2008}.

For vector-valued structural fields with multiple channels $\{\rho^{(m)}_s\}_{m=1}^M$, the full second-order spatial statistics are naturally expressed in terms of both autocorrelations and cross-correlations \cite{kaundinya_machine_2021, barry_voxelized_2023, barry_voxelized_2020}. We deliberately compute correlations of the square-rooted fields $\sqrt{\rho^{(m)}_s}$ so that the zero-vector value $f_0$ (the origin of the map at $r=0$) reduces exactly to the arithmetic mean of the field, thereby conserving the global average (e.g., the mean charge density or phase volume fraction) as an explicit, physically interpretable feature. The discrete two-point autocorrelation of channel $m$ and cross-correlation between channels $m$ and $n$ at separation $r$ are defined as:
\begin{equation}
f^{(mm)}_r = \frac{1}{|S|}\sum_{s\in S} \sqrt{\rho^{(m)}_s}\,\sqrt{\rho^{(m)}_{s+r}},
\quad
f^{(mn)}_r = \frac{1}{|S|}\sum_{s\in S} \sqrt{\rho^{(m)}_s}\,\sqrt{\rho^{(n)}_{s+r}},
\label{eq:multi_corr_real}
\end{equation}
for $m,n\in\{1,\dots,M\}$.
These quantities can be evaluated efficiently using FFTs (denoted by $\mathcal{F}$) as
\begin{equation}
f^{(mn)}_r = \frac{1}{|S|}\,\mathcal{F}^{-1}\!\big[
\mathcal{F}\!\big(\sqrt{\rho^{(m)}_s}\big)*
\mathcal{F}\!\big(\sqrt{\rho^{(n)}_s}\big)
\big],
\label{eq:multi_corr_fft}
\end{equation}
where the $m=n$ case reduces to the autocorrelation used elsewhere in this work \cite{cecen_versatile_2016, barry_voxelized_2023}. For notational simplicity, the remainder of this work focuses on the scalar case, with the understanding that the same FFT-based machinery applies channel-wise and to all auto- and cross-correlations.

While autocorrelations of a single scalar field already capture translationally invariant topology, cross-correlations between physically distinct channels (e.g., pseudo-density and elemental electronegativity, or occupancy and local void state) encode higher-order couplings that are otherwise only accessible to deep networks with many layers \cite{kaundinya_machine_2021, ray_lean_2025, barry_voxelized_2023}. The use of such cross-correlations as engineered features therefore offers a systematic and interpretable route to richer structure–property linkages without resorting to over-parameterized models \cite{kaundinya_machine_2021}.

This discrete summation inherently enforces translational invariance and naturally captures the periodicities, short-range order, and lattice distortions critical to the material's physical response. Standard Cartesian coordinate representations (e.g., XYZ, CIF, or POSCAR files)\cite{ong_python_2013} present significant limitations in this regard: atomic species are not natively interpolatable, selecting a definitive list of elemental descriptors introduces bias, and coordinate matrices are highly inefficient for representing complex defect populations. Furthermore, while raw electron charge density fields ($\rho_s$) are statistically homogeneous, a single DFT computation of a disordered molecular structure represents merely one instantiation out of an unknown number of possible instantiations that could be extracted for a given set of chemical species and their potential local configurations.

Conversely, two-point spatial correlations $f^{(mn)}_r$ capture the underlying structural motifs through the statistical moments of these stochastic fields. Because the correlation is defined on relative displacement vectors $r$, its output is indexed on a fixed, symmetric grid centered at $r=0$, providing a natural, unambiguous origin regardless of whether the input structure has any intrinsic reference point. This property makes the same mathematical machinery directly applicable to both perfectly periodic crystals and completely non-periodic finite molecules, without any arbitrary spatial alignment or padding convention. In practice, correlations for periodic crystals are evaluated with circular (periodic) FFT convolution on the supercell, whereas finite molecules are zero-padded to a common bounding box so that wrap-around contributions are excluded. The two conventions differ only in the treatment of the largest-$r$ vectors and leave the near-origin statistics that dominate the leading principal components unchanged. For defect-rich supercells containing hundreds of symmetry-inequivalent atoms, this fixed-origin, translationally invariant representation is particularly powerful, as direct coordinate-based descriptors become unwieldy and highly non-convex. By deterministically computing these spatial correlations via the FFT machinery of Eq.~\ref{eq:multi_corr_fft}, the computational burden of spatial pattern recognition is shifted away from the downstream machine learning algorithm, preventing the over-parameterization required by deep convolutional networks.

\begin{figure}[h]
    \centering
    \includegraphics[width=\linewidth]{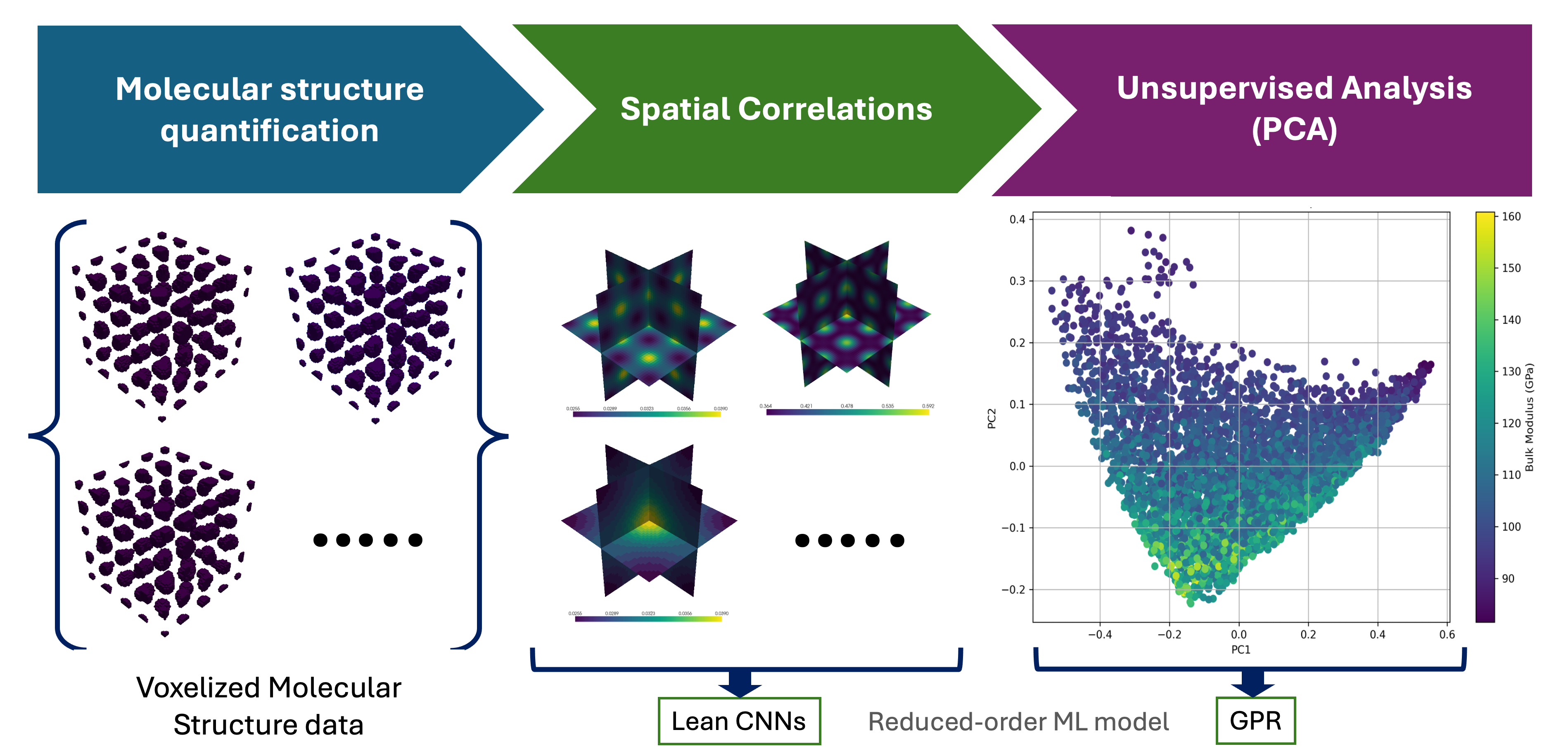}
    \caption{End-to-end spatial feature engineering workflow. Discrete atomic coordinates are first embedded into voxelized eigen–microstructure and pseudo-density fields on a standardized grid. Directionally resolved two-point auto- and cross-correlations are then computed via FFTs to obtain translationally invariant structural fingerprints. Finally, PCA projects these high-dimensional statistics onto a low-dimensional, convex manifold that serves as input to lean CNNs or GPR models.}
    \label{fig:pipeline}
\end{figure}

The high-dimensional nature of the computed two-point correlations necessitates a robust dimensionality reduction strategy (see Figure~\ref{fig:pipeline}) to construct a tractable input space for predictive modeling. Principal Component Analysis (PCA) \cite{mackiewicz_principal_1993} is systematically applied to the ensemble of two-point statistics, projecting the high-dimensional spatial data onto an orthogonal, low-dimensional feature space \cite{kalidindi_hierarchical_2015} that retains the vast majority of the structural variance.  This chemically agnostic, principal component manifold serves as a compact structural descriptor that seamlessly integrates with highly data-efficient machine learning algorithms, notably Gaussian Process Regression (GPR) \cite{rasmussen_gaussian_2006}. Operating within this reduced manifold not only provides crucial uncertainty quantification necessary for Bayesian active learning \cite{lookman_active_2019, balachandran_adaptive_2016, seyed_mahmoud_sequential_2026} but also isolates complex physical phenomena, distinguishing between global property dependencies driven by 3D volumetric distributions and localized structural effects.

In summary, the spatial feature engineering paradigm provides a practically deployable route for digitizing and interrogating complex molecular and crystalline structures for computational design. By combining voxelized discrete fields with two-point spatial correlations and principal component analysis, the framework yields a highly compressed, interpretable, and predictive structural manifold that can be used immediately with existing surrogate modeling tools. Consequently, it reduces the severe data dependencies of traditional deep learning and enables highly sample-efficient surrogate models for high-throughput screening workflows. For clarity, we organize the comparison around six operating regimes: (i) local energy/force learning, (ii) global property screening, (iii) chemically disordered systems, (iv) finite molecular conformations, (v) defect-sensitive localization, and (vi) active/inverse design. Spatial statistics are most advantageous for (ii)-(iv) and (vi); graph/equivariant models remain superior for (i) and (v) \cite{drautz_atomic_2019, batzner_e3-equivariant_2022}; and hybrid pipelines that feed physics-informed features into, or fine-tune, pretrained models are the most promising route for (ii)-(iv).

\subsubsection{Invariance, uniqueness, and degeneracy}

Two-point correlations defined on relative displacement $r$ are translationally invariant, and the voxelized field is permutation invariant by construction. They are \emph{not} automatically rotationally invariant: directionally resolved maps require a canonical orientation, spherical/rotational averaging, augmentation, or an equivariant/scattering treatment \cite{eickenberg_solid_2017} to attain rigid-motion invariance. Two-point statistics are also not a unique structural descriptor: distinct, non-congruent structures can share identical pair statistics (homometric degeneracy) \cite{patterson_homometric_1939, patterson_ambiguities_1944, gommes_microstructural_2012, cluff_quantifying_2026}. Multi-channel cross-correlations and higher-order statistics reduce, but do not eliminate, this ambiguity. The set of admissible two-point statistics is itself convex \cite{niezgoda_delineation_2008}, and because PCA is a linear projection, its image (the low-dimensional PCA manifold) inherits this convexity. Interpolation between valid states therefore remains within the admissible region, so the PCA representation supports continuous path-planning and inverse design, avoiding the pathological local minima of learned nonlinear latent spaces.

\subsection{Computational Infrastructure}
\subsubsection{Data Engineering}
The rapid generation of these spatial features relies heavily on advanced data engineering practices \cite{brough_materials_2017}. Transforming atomic coordinates into voxelized fields and subsequently computing spatial correlations is greatly accelerated by vectorized implementations of Fast Fourier Transform (FFT) algorithms on regular grids. Furthermore, the growing availability of open-source codebases, such as those provided for the Alloy Discovery \cite{ray_electronic_2026} and PFAS analyses \cite{ray_ml_2026} has significantly lowered the barrier to entry. These repositories offer standardized pipelines for data generation and curation, enabling rapid deployment of robust feature engineering workflows.

\begin{table}[t]
    \centering
    \footnotesize
    \caption{Reproducibility resources for the case studies. All repositories are open-source Jupyter/Python; consult each repository's LICENSE for terms.}
    \label{tab:repro}
    \begin{tabular}{p{2.8cm}p{2.9cm}p{3.0cm}}
        \toprule
        \textbf{Case study} & \textbf{Code repository} & \textbf{Key dependencies} \\
        \midrule
        Alloy discovery \cite{ray_electronic_2026} & \href{https://github.com/pranoy-ray/AlloyDiscovery}{github.com/pranoy-ray/AlloyDiscovery} & ase, pymatgen, NumPy/SciPy (FFT), gpytorch (GPR) \\
        PFAS screening \cite{ray_ml_2026} & \href{https://github.com/pranoy-ray/screenPFAS}{github.com/pranoy-ray/screenPFAS} & ase, pymatgen, RDKit, scikit-learn (GPR, RF), SHAP \\
        Electronic structure screening with Lean CNNs \cite{ray_lean_2025} & \href{https://github.com/pranoy-ray/LeanCNN}{github.com/pranoy-ray/LeanCNN} & ase, NumPy/SciPy (FFT), PyTorch/TensorFlow, pymatgen \\
        Core 2-pt statistics for multiscale applications \cite{brough_materials_2017} & \href{https://github.com/materialsinnovation/pymks}{github.com/materialsinnovation/pymks} & NumPy, SciPy, Dask \\
        \bottomrule
    \end{tabular}
\end{table}

Because all spatial statistics are computed deterministically via FFTs on standardized Eulerian grids, the computational complexity of the spatial feature generation scales fundamentally as $\mathcal{O}(S \log S)$, where $S$ is the total number of spatial voxels defining the bounding box of the molecular configuration. Furthermore, the computational cost scales quadratically with the number of physical attribute channels $M$ evaluated, $\mathcal{O}(M^2)$, as all permutations\footnote{in practice we only extract $\sim M$ cross-correlations, because there are a number of interdependencies between them.
} of cross-correlations are extracted. This represents a substantial algorithmic advantage over message-passing graph networks in the global-screening regime, whose per-pass cost scales approximately as $\mathcal{O}(L\,|E|\,d)$ for $L$ message-passing layers, $|E|$ edges, and hidden width $d$, with additional overhead for neighbor construction, cutoff policy, and batching. Importantly, the two-point-statistics cost is essentially invariant to the material class and is governed only by the grid discretization, since the voxel count and memory scale as $S \propto (L_{\text{box}}/h)^3$ with bounding-box size $L_{\text{box}}$ and grid spacing $h$. Consequently, the marginal cost of extracting features for any new molecular configuration is strictly bounded by the grid resolution rather than the number of atoms, ensuring that wall-times per voxelized simulation grid remain on the order of seconds on a single CPU core~\cite{cecen_versatile_2016, barry_voxelized_2023, barry_voxelized_2020, ray_lean_2025}. This stands in sharp contrast to graph-based or message-passing models that repeatedly execute expensive neighbor-aggregation kernels during both training and inference.

\subsubsection{ML/AI tools}
A profound advantage of spatial statistics–based models is their ability to be trained and deployed on minimal computational resources, in sharp contrast to computationally intensive graph neural networks (GNNs) and 3D CNNs that operate directly on raw atomic graphs \cite{xie_crystal_2018, choudhary_atomistic_2021, chen_graph_2019, reiser_graph_2022, kearnes_molecular_2016, gasteiger_gemnet_2024, duvenaud_convolutional_2015, choudhary_machine_2018, gurnani_polyg2g_2021}, or charge-density voxels \cite{zhao_predicting_2020, casey_prediction_2020, merchant_scaling_2023}. State-of-the-art GNNs often involve $10^6$–$10^8$ parameters, deep message-passing stacks, and batched neighbor searches, necessitating multi-GPU nodes and thousands of labeled structures to avoid overfitting; even inference latency can be prohibitive in active learning loops or screening campaigns \cite{ray_electronic_2026, barry_voxelized_2023} .

In contrast, models operating on deterministic spatial correlations completely bypass message passing, because the expensive step of learning spatial patterns is executed once, in closed form, by FFT-based feature engineering \cite{barry_voxelized_2023,ray_lean_2025,ray_ml_2026}. Reduced-order convolutional architectures built on these compressed manifolds eliminate fully connected layers and operate with $\mathcal{O}(10^4)$–$\mathcal{O}(10^5)$ parameters, making them comfortably trainable on a single commodity GPU \cite{ray_lean_2025}. Table~\ref{tab:models} summarizes these parameterization differences, and Table~\ref{tab:regimes} contrasts the two families across representative operating regimes. For Bayesian active learning workflows that rely on Gaussian Process Regression (GPR) in the PCA-reduced correlation space, all training and inference can be performed on a single CPU, enabling fully serial, laptop-scale campaigns that still achieve sub‑2\% errors with training sets as small as 10–30 DFT points \cite{barry_voxelized_2023, ray_electronic_2026}.

\begin{figure}[h]
    \centering
    \includegraphics[width=\linewidth]{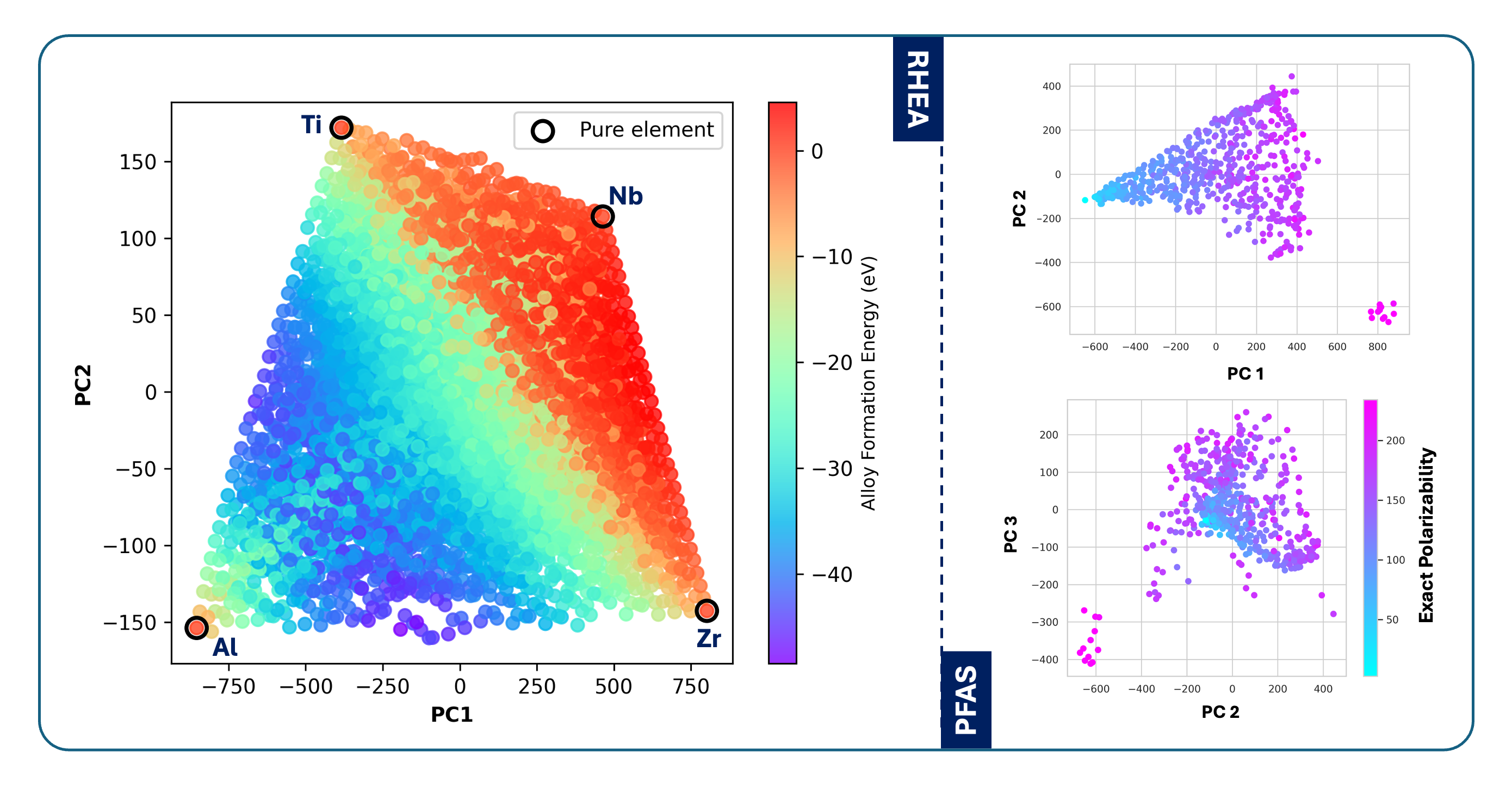}
    \caption{Broad applicability of the spatial statistics based feature engineering paradigms. The left panel (adapted from Ray et al., \textit{Digital Discovery 2026} \cite{ray_electronic_2026}) shows the Principal Component space of refractory high-entropy alloys, demonstrating the chemically agnostic grouping of elemental and multi-component systems. The right panel shows the correlation of the principal component scores with exact polarizability in PFAS, successfully isolating thermodynamic outliers and variations driven by the length of the perfluoroalkyl chain (adapted from Ray et al., \textit{Advanced Science 2026} \cite{ray_ml_2026}). Panels reproduced under the original publishers' reuse licenses.}
    \label{fig:applications}
\end{figure}

\begin{table}[h]
    \centering
    \caption{Comparison of parameterization and feature-generation cost across ML architectures. Parameter counts refer to \emph{trainable} parameters only; feature-generation cost is the FFT-based preprocessing per structure, scaling as $\mathcal{O}(S\log S)$ in the voxel count $S$.}
    \label{tab:models}
    \begin{tabular}{lccc}
        \toprule
        \textbf{Model Architecture} & \textbf{Feature Representation} & \textbf{Approx. Fittable Parameters} & \textbf{Feature-gen cost} \\
        \midrule
        Deep Graph Neural Networks & Node/Edge embeddings & $>10^{6}$ & Amortized in training \\
        Standard 3D CNNs & Raw Voxelized Densities & $>10^{6}$ & Voxelization only \\
        Lean CNNs & Spatial Correlations & $<10^{5}$ & FFT, $\mathcal{O}(S\log S)$ \\
        Gaussian Process Regression & PCA-reduced Correlations & Non-parametric & FFT + PCA \\
        \bottomrule
    \end{tabular}
\end{table}

\begin{table}[h]
    \centering
    \footnotesize
    \caption{Regime-based comparison of representation choices. Entries indicate typical, not universal, behavior; data needs depend on target property, sampling protocol, and training strategy (pre-training, transfer, $\Delta$-, or multi-fidelity learning can lower them \cite{ramakrishnan_big_2015, smith_approaching_2019}).}
    \label{tab:regimes}
    \begin{tabular}{p{2.3cm}p{2.6cm}p{1.6cm}p{1.4cm}p{1.6cm}p{2.4cm}}
        \toprule
        \textbf{Regime / task} & \textbf{Preferred representation} & \textbf{Typical labels} & \textbf{Compute} & \textbf{UQ / AL} & \textbf{Known failure modes} \\
        \midrule
        Local energy \& forces & SOAP/ACE, equivariant MLIP \cite{drautz_atomic_2019, batatia_mace_2022, deng_chgnet_2023} & $10^4$--$10^6$ & GPU & Ensemble/GP & Global targets; species combinatorics \\
        Global property screening & Two-point stats + lean ML \cite{barry_voxelized_2023, ray_electronic_2026} & $10$--$10^2$ & CPU/1 GPU & GPR (native) & Localized, bond-specific targets \\
        Chemically disordered systems & Two-point / cross-correlations \cite{ray_ml_2026, barry_voxelized_2023} & $10$--$10^2$ & CPU/1 GPU & GPR & Homometric degeneracy \cite{gommes_microstructural_2012} \\
        Finite molecular conformations & Void-state correlations, scattering \cite{ray_ml_2026, eickenberg_solid_2017} & $10^2$--$10^3$ & CPU/1 GPU & GPR & Orientation / boundary handling \\
        Defect-sensitive localization & Phase-preserving CNN, $n$-point \cite{kelly_recurrent_2021, muth_neighborhood_2023} & $10^3$+ & GPU & Bayesian NN & Rare-event tails smoothed by 2-pt stats \\
        Active / inverse design & PCA + GPR + acquisition \cite{lookman_active_2019, kalidindi_hierarchical_2015} & $10$--$10^2$ & CPU & Native & Non-uniqueness of target motifs \\
        \bottomrule
    \end{tabular}
\end{table}

\begin{figure}[t]
    \centering
    \includegraphics[width=0.85\linewidth]{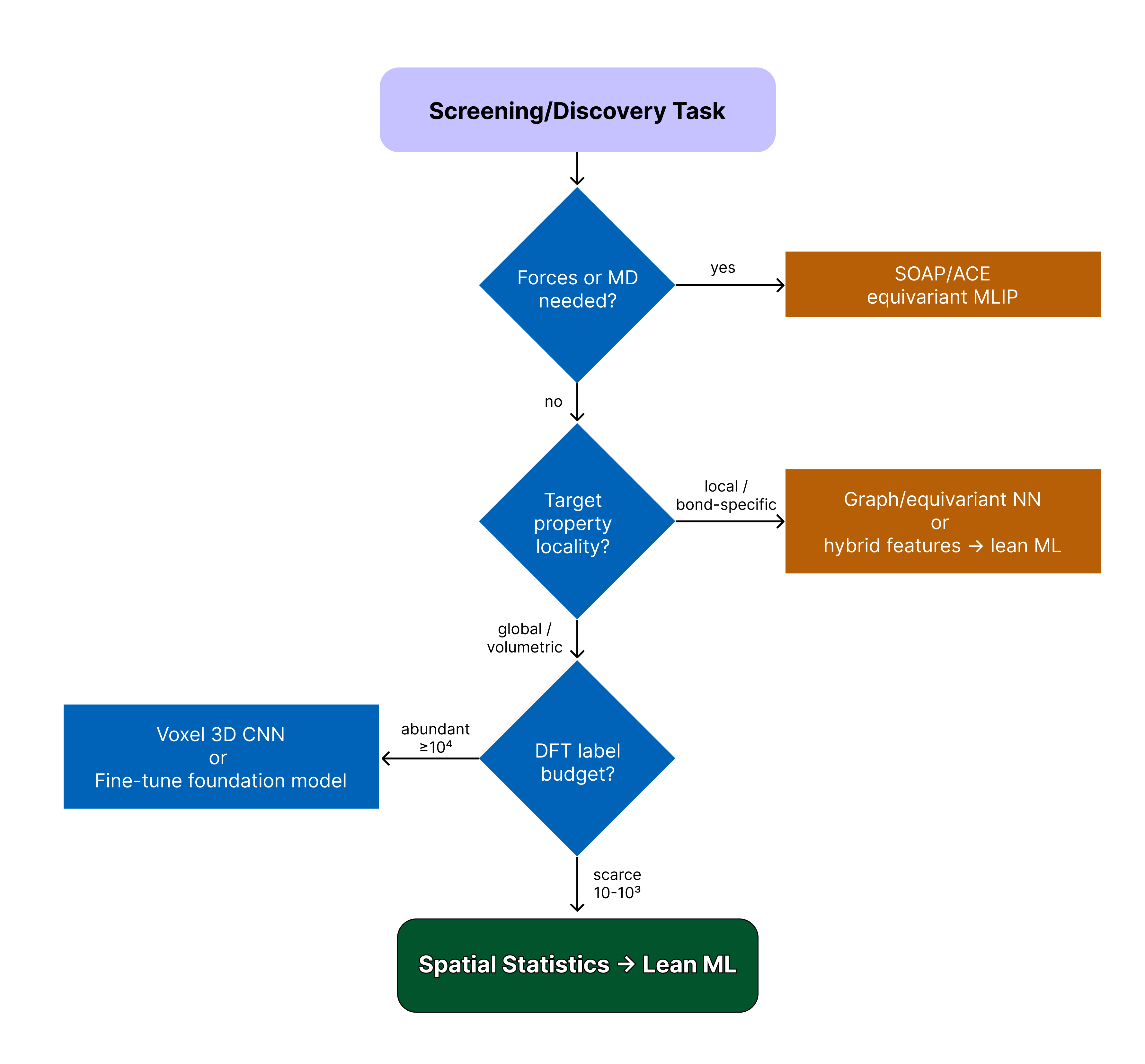}
    \caption{Decision guidance for representation choice. Starting from a screening task, the branch points are the need for forces/dynamics, the locality of the target property, and the DFT-label budget; the leaves recommend a representation family. Two-point statistics are the data-efficient default for global properties in the scarce-label regime, with disorder and chirality handled through cross- and void-state correlations. The companion Table~\ref{tab:regimes} lists the same regimes with typical data budgets, compute, and failure modes.}
    \label{fig:decision}
\end{figure}

\section{Applications}

As the application of spatial feature engineering transitions from perfect crystals to highly disordered structures and ultimately to isolated, non-periodic molecules, both the structural data and the corresponding physical phenomena become increasingly complex \cite{kalidindi_hierarchical_2015, barry_voxelized_2023, ray_ml_2026}. Perfect crystals exhibit exact translational symmetry and a small set of symmetry-inequivalent atomic environments; in this regime, a limited number of directionally resolved autocorrelation vectors is sufficient to resolve the periodic topology and elastic response \cite{barry_voxelized_2023, ray_lean_2025}. Disordered solid solutions \cite{zunger_special_1990} and glasses introduce local chemical fluctuations and strain fields that break perfect symmetry while preserving statistical homogeneity, demanding a broader spectrum of correlation vectors and sometimes cross-correlations between channels to capture the relevant disorder-induced motifs \cite{ray_ml_2026, barry_voxelized_2023, miracle_critical_2017, zheng_quick_2019}. This capability is particularly critical for the design of next-generation energy storage systems, such as high-capacity cation-disordered rock salt (DRX) battery cathodes \cite{urban_configurational_2014}. In these systems, the exact spatial distribution of redox-inactive elements and the resulting local cation fluctuations dictate the battery's voltage profile and long-term cycling stability - features that spatial correlations can capture and map directly to performance metrics where standard descriptors fail. Finally, isolated molecules abandon periodicity altogether and develop property-controlling features tied to global 3D conformation, steric shielding, and finite-size boundary effects, all of which are naturally encoded in volumetric fields (e.g., pseudo-density or void-state: see Figure~\ref{fig:applications}) but are often poorly represented in graph-only descriptors \cite{ray_ml_2026, gomberg_extracting_2017}.

Beyond crystalline solids, the spatial feature engineering paradigm provides a robust mechanism for capturing the conformational complexity of organic molecules, which is often obfuscated by standard topological graph representations. In the realm of practical drug discovery, this framework offers a distinct advantage for characterizing complex binding pockets and intrinsically disordered proteins (IDPs). Because IDPs exist as highly dynamic conformational ensembles rather than rigid 3D structures, traditional virtual screening and point-cloud methods frequently fail to capture their behavior \cite{uversky_how_2025, ravarani_highthroughput_2018}. By converting these fluctuating ensembles into spatially correlated pseudo-density or voxelized fields, researchers can extract stable, low-dimensional statistical signatures representing the protein's true functional phase space. This opens a direct route to structure-function linkages for disordered proteins that are otherwise intractable for rigid-structure docking and graph-based screening methods. 

\begin{figure}[h]
    \centering
    \includegraphics[width=0.8\linewidth]{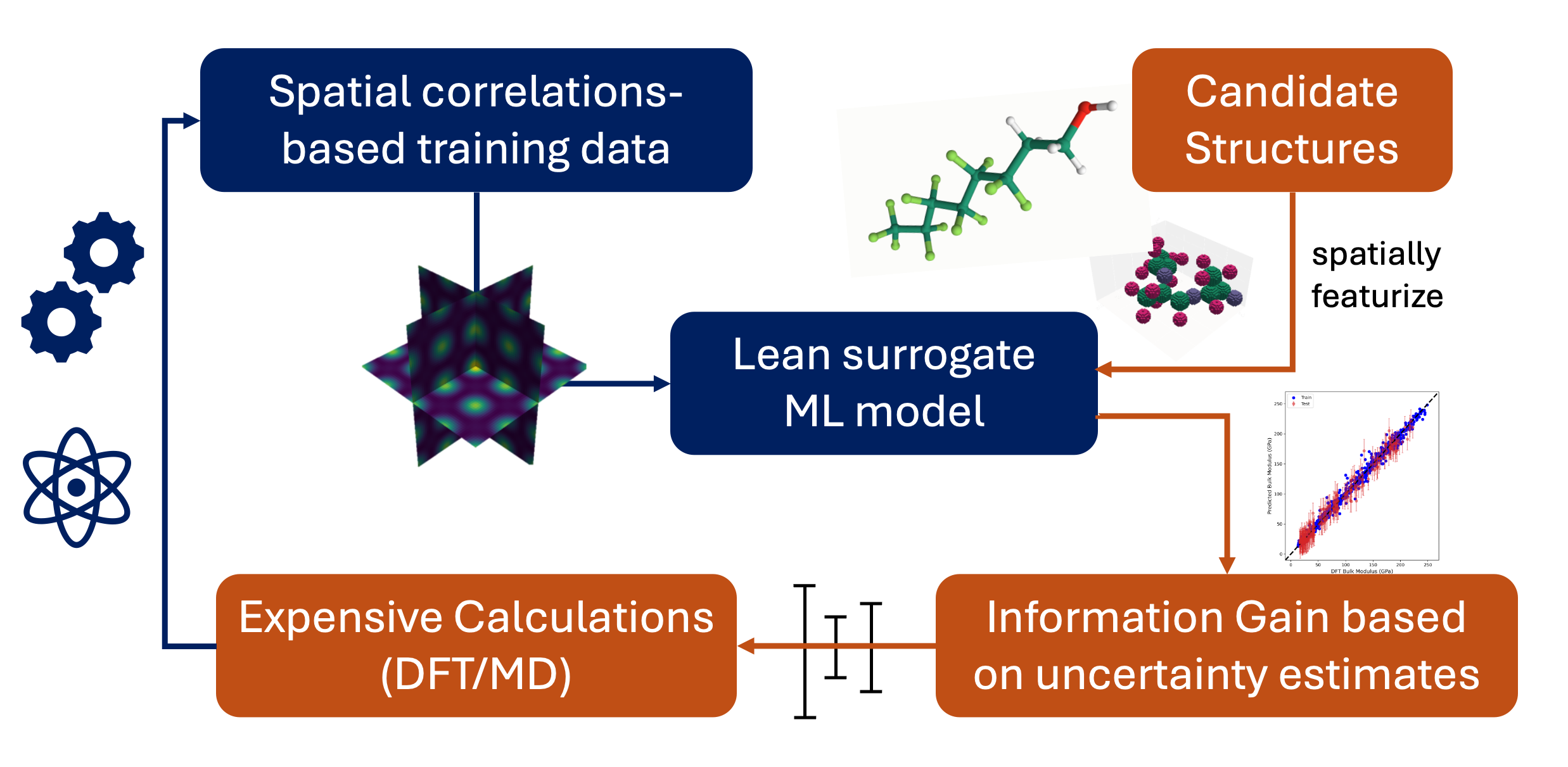}
    \caption{Integration into autonomous discovery pipelines. Conceptual closed-loop workflow in which voxelization, FFT-based spatial statistics, and PCA feed into lean surrogates that provide both property predictions and uncertainty estimates. These outputs then inform Bayesian experiment design, guiding expensive DFT or MD calculations only to the most informative candidate structures.}
    \label{fig:future}
\end{figure}

Beyond static configurations, the same pair-correlation-plus-PCA description has been applied directly to molecular-dynamics data: Gomberg et al. used a specialized pair-correlation-function descriptor, reduced to its leading principal components, to learn structure--property linkages (solute segregation) at grain boundaries from molecular-mechanics simulations \cite{gomberg_extracting_2017, kalidindi_application_2015}. Similarly, a critical environmental application (see Table~\ref{tab:applications}) is the characterization and remediation of per- and polyfluoroalkyl substances (PFAS), where persistence is fundamentally linked to rigid, helically packed backbones and the associated organization of molecular void space. Whether deployed on density functional theory (DFT) datasets generated via VASP\cite{kresse_efficiency_1996, kresse_ultrasoft_1999, kresse_efficient_1996} or ORCA\cite{neese_orca_2012, neese_software_2022}, or molecular dynamics (MD) trajectories simulated via LAMMPS, the extraction of spatial correlations enables high-fidelity property prediction \cite{ramprasad_machine_2017, himanen_data-driven_2019}. Figure~\ref{fig:future} summarizes how these components can be integrated into autonomous, closed-loop materials discovery pipelines driven by active learning. 

Contemporary Bayesian active learning frameworks \cite{lookman_active_2019, balachandran_adaptive_2016}, which incorporate dynamic gating mechanisms to prioritize the most informative regions of chemical space, have demonstrated dramatic reductions in the number of experiments required for targeted materials discovery  \cite{khatamsaz_bayesian_2023, ray_refining_2025, seyed_mahmoud_sequential_2026, buzzy_active_2025}. By operating on the low-dimensional, mathematically convex spatial manifolds generated by PCA, the proposed framework complements these advanced closed-loop systems. The convex nature of the latent space ensures that Bayesian acquisition functions (e.g., Expected Improvement, Upper Confidence Bound) can be optimized continuously without succumbing to the pathological local minima that plague the highly non-convex latent spaces of deep generative models. This efficiently solves the "one-to-many" problem inherent in inverse molecular design, where multiple disparate structural motifs may satisfy the target property \cite{noauthor_ai-accelerated_nodate}. Furthermore, because the mathematics of spatial statistics are inherently scale-agnostic, these correlation representations can be integrated alongside emerging Foundation Models for macroscopic structures (e.g., the PolyMicros framework \cite{buzzy_polymicros_2025}) to establish rigorous multiscale linkages from atoms to polycrystalline grains.

\begin{table*}[t]
    \centering
    \footnotesize
    \caption{Evidence for the sample-efficiency claims. Each row is a demonstrated case study with its representation, surrogate, data budget, evaluation setting, and reported error. Validation is $k$-fold cross-validation unless noted; ``extrap.'' denotes hold-out on chemistries absent from training. Code links are given in Table~\ref{tab:repro}.}
    \label{tab:applications}
    \begin{tabular}{p{2.2cm}p{2.3cm}p{2.9cm}p{1.5cm}p{1.3cm}p{2.6cm}}
        \toprule
        \textbf{Material class} & \textbf{Target property} & \textbf{Representation + surrogate} & \textbf{Train / total} & \textbf{Setting} & \textbf{Reported error} \\
        \midrule
        Perfect crystals \cite{kaundinya_machine_2021, ray_lean_2025} & Formation / effective properties & 2-pt correlations + lean CNN ($<$81K params) & subset / 1410 & interp. & low normalized error (see ref.) \\
        Refractory HEAs \cite{ray_electronic_2026} & Bulk modulus & non-interacting density, 2-pt + PCA + GPR (Bayesian AL) & 10--30 & extrap. & NMAE $<2\%$ \\
        Small molecules (PFAS) \cite{ray_ml_2026} & Polarizability; enthalpy & 2-pt + PCA + GPR & per ref. & interp. & $R^2\approx0.92$; $0.97$ \\
        Small molecules (PFAS) \cite{ray_ml_2026} & Bond dissociation energy & graph features + random forest & per ref. & interp. & $R^2\approx0.87$ \\
        Grain boundaries \cite{gomberg_extracting_2017, kalidindi_application_2015} & Solute segregation (structure--property) & pair-correlation function + PCA & per ref. & interp. & see ref. \\
        \bottomrule
    \end{tabular}
\end{table*}

In the specific case of PFAS, recent ML workflows have demonstrated that predicting degradation-relevant properties requires fundamentally different featurization schemes depending on the underlying physics of the target property \cite{ray_ml_2026}. Global, volumetric properties such as polarizability and thermodynamic enthalpy are highly dependent on long-range 3D electron density patterns and are successfully predicted using the spatial correlation and PCA framework ($R^2 \approx 0.92$ to $0.97$) \cite{ray_ml_2026}. Conversely, highly localized, bond-specific properties such as Bond Dissociation Energies (BDEs) which govern reaction kinetics remain better suited for topological graph-based representations \cite{ray_ml_2026}.

Crucially, regarding conformational geometry, spatial statistics have been used to define chemistry-agnostic helicity descriptors by computing \emph{directionally resolved} autocorrelations of the molecular void-state channel (and void--atom cross-correlations) rather than of the atoms themselves \cite{ray_ml_2026}. Because these maps are not rotationally averaged, they retain the handedness of the configuration, and their projection onto the leading principal components yields a continuous measure that separates chiral, helical perfluoroalkyl chains from planar hydrogenated analogues \cite{ray_ml_2026}. We emphasize that this is an empirical separation demonstrated on the tested systems rather than a general chirality guarantee; it nonetheless contrasts with the $E(3)$-reflection limitation of standard equivariant models \cite{dumitrescu_e3-equivariant_2025}. This example illustrates how spatial correlations can simultaneously encode local bonding motifs and long-range chiral conformational order in non-periodic volumes. In such systems, cross-correlations between atomic and void channels, or between pseudo-density and chemically weighted occupancy fields, offer a rich source of global information that is extremely challenging to learn robustly with standard point-cloud or message-passing models at small data scales \cite{kaundinya_machine_2021, ray_ml_2026}.

Taken together, the preferred representation in each setting (summarized in Figure~\ref{fig:decision} and Table~\ref{tab:regimes}) follows from the data budget, property locality, disorder level, periodicity, force requirements, chirality sensitivity, uncertainty quantification, and inverse-design needs.

\section{Conclusions}

Data-driven surrogate models have enabled the rapid screening, prognosis, and design of complex materials and molecular systems. This review outlines both the conceptual foundation and the mathematical framework required to transition materials informatics away from over-parameterized deep learning architectures toward highly sample-efficient models. Spatial feature engineering provides a physically rigorous, translationally invariant representation of the underlying structural and electronic topology. The proposed framework consists of a molecular representation based on voxelized continuous fields and two-point spatial correlations, a dimensionality reduction pipeline utilizing Principal Component Analysis (PCA), and a property prediction architecture centered around reduced-order convolutional neural networks and Bayesian-driven Gaussian Process Regression. Together, these foundational elements offer concrete opportunities for the extrapolation and exploration of complex chemical spaces, bridging the critical gap between expensive quantum mechanical accuracy and the accelerated demands of autonomous materials discovery.

Taken together, the emerging literature on voxelized fields, spatial correlations, and pseudo-density–based manifolds demonstrates that much of what deep GNNs and large 3D CNNs attempt to learn implicitly can be computed deterministically, interpreted physically, and compressed into low-dimensional, convex feature spaces \cite{kalidindi_hierarchical_2015, kaundinya_machine_2021, barry_voxelized_2023, lei_universal_2022, ray_lean_2025, barry_voxelized_2020, ray_ml_2026, ray_electronic_2026}. This review argues that for the vast majority of realistic screening and discovery workflows, where DFT labels are scarce and extrapolation across chemistries is essential, physics-informed spatial feature engineering coupled to lean surrogates is not merely an alternative to graph-based deep learning, but a competitive, resource-light default in data-scarce, extrapolative regimes.

\textit{Limitations and failure modes:} Despite these advantages, 2-point spatial statistics–based representations are not a panacea. While they excel at capturing volume-averaged, defect-insensitive bulk properties (homogenization), they inherently smooth over the rare, localized topological anomalies responsible for catastrophic material failure. These limitations and their mitigations are summarized in Table~\ref{tab:failure}. Predicting highly localized, defect-sensitive phenomena such as bond-specific cleavage energies, damage initiation, or fatigue hot-spots requires moving beyond pairwise interactions. These phenomena are governed by extreme value distributions and weakest-link theories, necessitating the capture of complex, high-order spatial correlations (e.g., $n$-point statistics)~\cite{venkatraman_reduced-order_2020, muth_neighborhood_2023}. While extracting these high-order statistics deterministically via FFTs suffers from the curse of dimensionality, recent breakthroughs in generative materials informatics offer an elegant solution. Advanced frameworks utilizing Local-Global Decompositions (LGD)~\cite{robertson_localglobal_2023} and score-based denoising diffusion models~\cite{ buzzy_statistically_2024} have successfully demonstrated the ability to condition global topologies on 2-point statistics while utilizing neural diffusion refinement to implicitly embed the crucial higher-order spatial correlations necessary for realistic, defect-aware microstructures. Continued exploration of these generative high-order embeddings, coupled with the rigorous full-field mapping capabilities of recurrent localization networks operating on raw voxelized fields~\cite{kelly_recurrent_2021}, represents the natural next step in establishing a truly generalizable, multiscale materials knowledge system. Accordingly, we regard spatial-statistics-only pipelines and hybrid spatial-statistics/deep-learning approaches (physics-informed features feeding, or fine-tuning, pretrained and foundation models) as complementary rather than competing directions, each preferable in the operating regimes identified in Table~\ref{tab:regimes}.

\begin{table}[h]
    \centering
    \footnotesize
    \caption{Failure modes of two-point spatial-statistics representations and mitigation strategies.}
    \label{tab:failure}
    \begin{tabular}{p{3.0cm}p{4.2cm}p{4.4cm}}
        \toprule
        \textbf{Failure mode} & \textbf{Mechanism} & \textbf{Mitigation} \\
        \midrule
        (a) Homometric degeneracy & Distinct structures share identical pair statistics \cite{patterson_homometric_1939, gommes_microstructural_2012} & Multi-channel cross-correlations; higher-order ($n$-point) statistics \\
        (b) Resolution / smearing sensitivity & Voxel size and smearing kernel bias the field & Convergence study over grid spacing and kernel width \\
        (c) Aliasing / discretization error & Under-sampling of sharp features & Nyquist checks; anti-aliasing before FFT \\
        (d) Field / channel-design dependence & Choice of channels shapes the descriptor & Report channel set; ablate channels \\
        (e) Localized reaction coordinates & Bond-breaking averaged out by pair statistics & Pair with graph / local descriptors \\
        (f) Channel / cross-correlation scaling & Cost grows as $\mathcal{O}(M^2)$ in channels & Prune interdependent cross-correlations ($\sim M$) \\
        (g) Finite-molecule orientation / boundary & Grid orientation and padding affect maps & Canonical orientation; fixed bounding box / zero-pad \\
        (h) PCA-basis domain shift & Basis fit on one chemistry misfits another & Re-fit or pool PCA basis across chemistries \\
        \bottomrule
    \end{tabular}
\end{table}

\section{Author contributions}

\textbf{Pranoy Ray}: Conceptualization, Methodology, Software, Validation, Formal analysis, Investigation, Data curation, Writing – original draft, Writing – review \& editing, Visualization. \textbf{Surya R. Kalidindi}: Conceptualization, Methodology, Validation, Resources, Writing – review \& editing, Supervision, Project administration, Funding acquisition. 

\section{Acknowledgements}

PR and SK acknowledge support from NSF DMREF Award 2119640. PR also acknowledges support received from the Novelis Graduate Scholarship and the William H. Glenn Sr. Fellowship.

\section{Conflicts of interest}
There are no conflicts to declare.

\printendnotes

% Submissions are not required to reflect the precise reference formatting of the journal (use of italics, bold etc.), however it is important that all key elements of each reference are included.
\bibliography{biblio}

%\begin{biography}[example-image-1x1]{A.~One}
%Please check with the journal's author guidelines whether author biographies are required. They are usually only included for review-type articles, and typically require photos and brief biographies (up to 75 words) for each author.
%\bigskip
%\bigskip
%\end{biography}

%{Please check the journal's author guildines for whether a graphical abstract, key points, new findings, or other items are required for display in the Table of Contents.}

\end{document}